\newcommand{\eps}{{\varepsilon}}

\def\ie		{{\it i.e.,~}}

\def\vs		{{\it vs.~}}

\documentclass[twocolumn,showpacs,preprintnumbers,amsmath,amssymb]{revtex4}

\usepackage{graphicx}
\usepackage{dcolumn}
\usepackage{bm}

\begin{document}

\preprint{APS/123-QED}

\title{Shadowing high-dimensional Hamiltonian systems: the gravitational $n$-body problem}

\author{Wayne B. Hayes}
\email{wayne@cs.toronto.edu}
\affiliation{%
Department of Computer Science,
University of Toronto,
Toronto, Ontario, M5S 3G4
Canada
}%


\date{\today}

\begin{abstract}
A {\it shadow} is an exact solution to a chaotic system of equations
that remains close to a numerically computed solution for a long time,
ending in a {\it glitch}.  We study the distribution of shadow
durations at low dimension and how shadow durations scale as dimension
increases up to 300 in a slightly simplified gravitational $n$-body
system.  We find that ``softened'' systems are shadowable for many tens
of crossing times even for large $n$, while in an ``unsoftened'' system
each particle encounters glitches independently as a Poisson process,
giving shadow durations that scale as $1/n$.
\end{abstract}

\pacs{02.60.-x, 02.70.-c, 05.45.-a, 05.45.Jn, 95.75.P, 05.40.-a, 98.10.+z
}
\maketitle


The behaviour of dynamical systems is often studied using
numerical techniques.  A source of error in such studies
arises because dynamical systems often display {\it
sensitive dependence on initial conditions}: two solutions whose
initial conditions differ by an arbitrarily small amount generally
diverge exponentially away from each other.
Since numerical methods introduce errors, it is virtually
guaranteed that a numerically computed solution diverges exponentially
away from the exact solution with the same initial conditions.  This
remains true even if integrals of motion such as energy and momentum
are conserved to arbitrary precision.  Although this fact is widely known,
its impact is not well understood.

Fortunately, most studies of dynamical systems do not aim to
predict the precise evolution of a particular choice of initial
conditions.  Instead, the dynamics of the system is {\em sampled}
in order to study its general behaviour.  In such cases, we typically
choose initial conditions from a random distribution and would be
happy if our numerical solution exhibited behaviour typical
of {\em any} valid choice of initial conditions from our
distribution.  In particular, we may be satisfied if our
numerical solution closely follows some exact solution whose initial
conditions are close to those that we chose.

The study of {\em shadowing} provides just such a property: a {\em
shadow} is an exact solution to a given set of equations that remains
close to a numerically computed solution of the same set of equations
for a nontrivial duration of time.  Although not all numerical
simulations are likely to be shadowable
\cite{DawsonGrebogiSauerYorke94,SauerGrebogiYorke97,Sauer02},
the existence of a shadow is a strong property: it asserts that a
numerical solution can be viewed as an {\em experimental observation}
of an exact solution.  As such, within the ``observational'' error, the
dynamics observed in a numerical solution that has a shadow represent
the dynamics of an exact solution.  There are only two remaining
questions (both beyond the scope of this paper): (1) Whether
the mathematical model being simulated accurately reflects the system
being studied.  This is certainly {\em not} the case for systems such
as the weather (and thus shadowing is probably an inappropriate measure
of error), but can be assumed to be the case for others, such as the
unsoftened gravitational $n$-body problem.  (2) Whether shadows
are typical of exact solutions chosen at random.
Simple examples exist of shadows that are atypical
\cite{FarmerSidorowich91,FryskaZohdy92,CorlessWhatGood94},
although it seems unlikely that atypical shadows
are common --- lest the numerical solutions we compute would be
commonly atypical as well \footnote{This question is not unique to
shadowing.  For example, even though a symplectic integrator applied
to a Hamiltonian problem exactly solves a nearby Hamiltonian problem
\cite{WisdomHolman92}, we could ask if the exactly solved problem is
typical of nearby Hamiltonian problems of interest.}.

Shadows of ``noisy'' trajectories were first proved to exist in {\it
hyperbolic} systems \cite{Anosov,Bowen}.  Such systems possess a phase
space which is spanned locally by an independent set of stable and
unstable directions that remain consistent under the evolution of the
system.  Few chaotic dynamical systems are globally hyperbolic, but for
those that are nearly hyperbolic over finite time intervals, the
existence of finite-duration shadows can often be inferred
\cite{GHYS,QT} or rigorously proved \cite{GHYS,CoomesKocakPalmer95a,HayesPhD}.
If a shadow is viewed as a measure of error of a numerical solution,
then the relevent measures are the phase-space distance between
corresponding points on the ``noisy'' and exact trajectories (smaller
is better), and the duration over which they remain
close together (longer is better).  Generally, the smaller the local
error in the trajectory, and the more hyperbolic the trajectory, the
closer and longer the shadow \cite{QT}. In
\cite{DawsonGrebogiSauerYorke94,SauerGrebogiYorke97,Sauer02},
scaling laws were developed for the expected
duration of shadows, based upon the distance from zero of finite-time
Lyapunov exponents of the system.  If an exponent fluctuates about zero
over a trajectory, it represents a tangential direction that is
uncertain between stable and unstable.  This effect, also called
unstable dimension variability \cite{LaiLernerWiliamsGrebogi99},
is the cause of unshadowability in chaotic systems.  If a numerical
trajectory is unshadowable, then it is possible (though far from
certain) that statistical quantities associated with the numerical
trajectory can have significant bias \cite{FryskaZohdy92,CorlessWhatGood94,Sauer02}.

To date, most studies of shadowability of chaotic systems have focussed
on maps or ODEs with only a few dimensions, and concern has been expressed
that fluctuating Lyapunov exponents may be common in high-dimensional systems
\cite{DawsonGrebogiSauerYorke94,SauerGrebogiYorke97,Sauer02}.
Furthermore, since Hamiltonian systems are not globally hyperbolic,
there is no reason to expect them to be easily shadowable.  In this
paper, we demonstrate that trajectories of at least one commonly integrated
continuous Hamiltonian system (the softened gravitational $n$-body problem)
are shadowable for long times with as many as 150 phase-space dimensions
(25 particles).
We also demonstrate that, for the purpose of shadowing, a large $n$-body
system can be viewed as a superposition of many small systems, so that the
distribution of shadow durations for a high-dimensional $n$-body system can
be approximated using the distribution of shadow durations of many
low-dimensional systems.

The trajectories that
we will attempt to shadow belong to a slightly simplified gravitational
$n$-body problem in which there are $N$ total particles, only $M$ of
which move.  We do this because the shadowing algorithm we use
takes time $O(M^3)$
and we want to simulate a large system with a complex potential while
keeping the time to compute a shadow tractable.
Each moving particle interacts both with fixed
particles and with other moving particles via Newton's gravitational force
law,
$$F_{ij} = -\frac{G m_i m_j}{r^2_{ij} + \eps^2},$$
where $F_{ij}$ is the force on particle $i$ from particle $j$, $m_i$
and $m_j$ are their masses, $r_{ij}$ is the distance between them, and
$\eps$ is the {\it gravitational softening parameter} which, if
non-zero, artificially smoothens the gravitational potential in order
to approximately emulate a system with more particles than are actually
present and to avoid the singularity at
$r_{ij}=0$.  We use normalized units \cite{HeggieMathieuNoXRef} in which each
particle has mass $1/N$, the system has diameter of order unity, and the
{\it crossing time} (the average time for a particle to cross the system
from one side to the other) is of order unity.
We use a variable-order, variable timestep integrator \cite{LSODE} for
all integrations.  We generate
noisy trajectories with local errors of about $10^{-5}$ per crossing time.
To find shadows, we use an algorithm described in \cite{QT},
optimized to run between two and three orders of magnitude faster.
Called {\it iterative refinement}, we use the same integrator as the
noisy trajectory with tighter tolerance to estimate the full phase-space
vector of local errors of the noisy trajectory, and then use a Newton-like
correction to refine the trajectory until it has local errors as small
as possible.  For simple systems, the errors of the refined trajectory
can be as small as the machine precision ($10^{-16}$), but the minimum
local error acheivable with refinement increases as the number of
dimensions increases, due to numerical errors in
computing the Newton corrections.
A trajectory produced by refinement is
called a {\it numerical shadow}.  The existence of a numerical shadow
is expected to indicate the existence of an exact shadow of comparable
duration \cite{QT}.

\begin{figure}
\includegraphics[width=8cm,height=5cm]{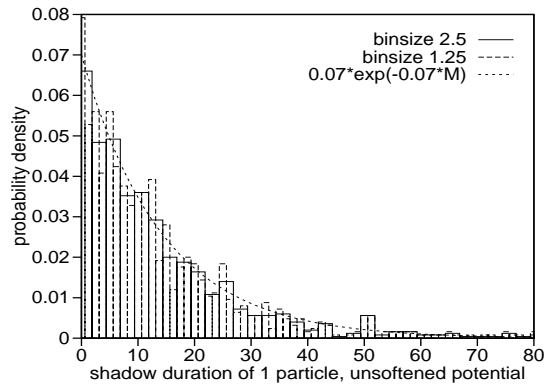}
\caption{Histogram of shadow durations 
of 1000 unsoftened gravitational $n$-body systems.
Each system has 99 fixed particles and one moving particle.
Noisy orbits have local error of about $10^{-5}$ per crossing time;
numerical shadows were required to have a maximum local error no bigger
than $10^{-14}$.  The horizontal axis is in crossing times; the
vertical axis is the measured probability density.  The distribution
fits an exponential curve with a mean glitch rate of about 0.07
per crossing time, indicating that the moving particle
encounters glitches as a Poisson process in an unsoftend system.
}
\label{fig:soft0-M1-expo-1e-14}
\end{figure}

A system with parameters $N=100,M=1,\eps=0$ was first shadowed by
\cite{QT}, who found that the single particle can be shadowed for
a few tens of crossing times, and that {\it glitches} (the point
beyond which a shadow cannot be found) tend to occur more frequently
during close approaches with other particles.  Combined with the fact
that close approaches occur as a stochastic process \cite{SpitzerGlob},
we hypothesized that glitches also occur as a stochastic process.
Fig.~\ref{fig:soft0-M1-expo-1e-14} shows a histogram of shadow
durations for 1000 $M=1$ systems.  The distribution is well-fit by an
exponential curve, suggesting that glitches are encountered
as a Poisson process in an unsoftened system.  This is moderately
surprising because it says that the number of crossing times that have
occured in the past has little influence on the occurance of a glitch
in the future, and thus there is in principle no natural upper limit
on the duration of a shadow.

\begin{figure}
\includegraphics[width=8cm,height=5cm]{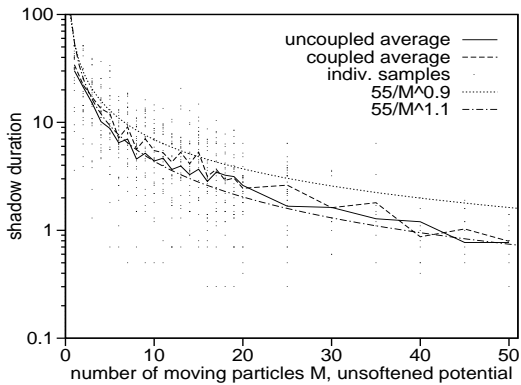}
\caption{ Average shadow durations in crossing times of an unsoftened
gravitational $n$-body system in which there are 100 particles, $M$ of
which move, as a function of $M$.  The noisy orbits have local error of
$10^{-5}$ per crossing time; each numerical shadow was required to have
a maximum error of $10^{-12}$.
The points represent sample shadow
durations, 30 samples each for $M=1,2,\ldots,19,20,25$ and 10 samples
each for $M=30,35,\ldots,50$.  The ``coupled average'' line joins their
averages, while $55/M^{0.9}$ and $55/M^{1.1}$ are plotted for comparison.
The ``coupled average'' line is statistically indistinguishable from the
``uncoupled average'' one in which the
gravitational interaction between moving particles is deleted, which
both validates the robustness of our shadowing algorithm for large $M$,
and suggests that even coupled particles encounter glitches independently
of one another.
}
\label{fig:soft0-coupled-vs-M}
\end{figure}

The robustness of our shadowing algorithm with increasing problem
dimension was tested in two ways.
First, we contrived
a slightly nonlinear Hamiltonian system designed to be easily shadowable 
(\ie almost hyperbolic), and successfully shadowed it for 20--50 Lyapunov
times on average while the dimension was increased from 2 to 180.
Second, we searched for shadows of gravitational systems similar
to the above consisting of $99+M$ particles
in which the $M$ moving particles interact only with the 99 fixed
particles.  Such a system is equivalent to superimposing $M$ uncoupled
single-particle systems, and it will encounter glitches as a Poisson
process with an aggregate rate $M$ times that of a single-particle system.
This results in average shadow durations that scale as $1/M$.  This was
in fact observed in our simulations, giving us confidence that our shadowing
algorithm works as well for large $M$ as it does for small $M$.

Fig.~\ref{fig:soft0-coupled-vs-M} demonstrates that the scaling is
{\em still} $1/M$ even if the $M$ moving particles interact.  This is
moderately surprising because it suggests that although particles interact
in motion, they do not interact in glitching.  A possible explanation is
that if $1<M\ll N$, then the coupling between the $M$ moving particles is
weak on average and we can still view the system as the superposition of
$M$ single-particle systems\footnote{The weak coupling assumption is broken
when a close approach between two moving particles occurs, but since it
is likely to cause a glitch anyway, and our argument need only apply until
a glitch occurs, the assumption holds during the time interval we need it
to hold.}.  Furthermore, we
note that the cross-section for close approaches is not altered as $M$
increases (which simply changes fixed particles into moving ones), so
the argument still holds independent of whether $M\ll N$.

\begin{figure}
\includegraphics[width=8cm,height=5cm]{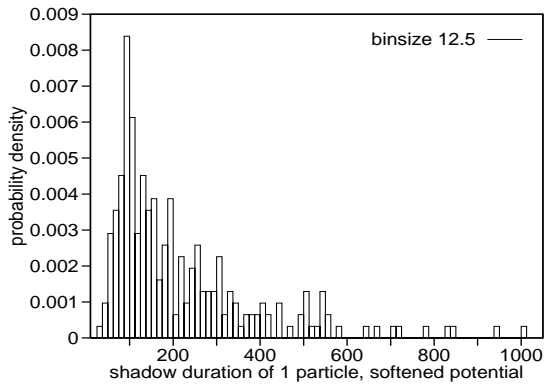}
\caption{Histogram of shadow durations of 250 systems identical to
those in Fig.~\ref{fig:soft0-M1-expo-1e-14} except with
softening $\eps=0.1$.  Note that, in contrast to
Fig.~\ref{fig:soft0-M1-expo-1e-14}, the $x$ axis extends to 1000 crossing
times, and that the histogram height is zero near a shadow duration of
zero, meaning that {\em no} particles undergo glitches until several
crossing times have occured; in fact, of the 250 systems sampled,
the shortest shadow lasts 37 crossing times.
}
\label{fig:soft-M1-hist}
\end{figure}

\begin{figure}
\includegraphics[width=8cm,height=5cm]{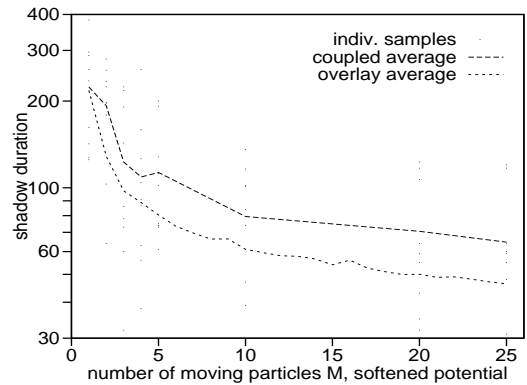}
\caption{
Average shadow durations in crossing times of a system with softening
parameter $\eps=0.1$ as a function of $M$.  The points represent sample
shadow durations, 9 samples each for $M=1$--$5,10,20,25$, while the
``coupled average'' line joins their averages.  Following the
assumption that particles encounter glitches independently of one
another, the ``overlay average'' is artificially constructed for each
$M=1,\ldots, 25$ by superimposing $M$ samples chosen at random from
Fig.~\ref{fig:soft-M1-hist} and taking the minimum shadow
duration of those samples.  The resulting graph resembles the ``coupled
average'' graph reasonably well, except for the surprising effect of
{\em under}-estimating shadow duration of the real system by about
(20$\pm$10)\%.
}
\label{fig:soft.1-coupled-vs-M}
\end{figure}

With non-zero softening, \cite{QT} found that shadow durations
increased significantly, while the correllation between glitches and
close approaches decreased.  Fig.~\ref{fig:soft-M1-hist} shows the
distribution of shadow durations for 250 $M=1$ systems in which
softening has been set to $\eps=0.1$, which is approximately half
the average inter-particle separation.  The differences from
Fig.~\ref{fig:soft0-M1-expo-1e-14} are quite marked.  First, the distribution
is peaked near 100 crossing times and has a long tail going out to
hundreds of crossing times.  Second, even
though the local errors of the noisy and shadow orbits are identical to
those from Fig.~\ref{fig:soft0-M1-expo-1e-14}, the average shadow length
has increased by more than an order of magnitude from 14 to 218.  Although
this is roughly equivalent to the increase in the Lyapunov timescale
\cite{QT}, the distribution is far from exponential.   In fact, the most
striking difference from Fig.~\ref{fig:soft0-M1-expo-1e-14} is that the
distribution has a vanishingly small density near zero shadow duration,
in striking contrast to a Poisson process.  In other words, {\em virtually no
particles undergo glitches until several tens of crossing times have
occurred}.  If this remains true even for $M>1$, then shadowing of
softened gravitational systems would be feasible even for large $M$,
because the trajectories of all particles in the simulation would
remain valid for many crossing times.
This question is addressed in Fig.~\ref{fig:soft.1-coupled-vs-M},
where we plot the average shadow duration for softened systems as
a function of $M$, along with the shadow duration predicted by
superimposing $M$ single-particle systems.  We see that although
the duration of shadows for coupled systems decreases as $M$ increases,
they decrease {\em much} more slowly than $1/M$, and appear to be
levelling off at about 50 crossing times.  This is consistent
with superimposing single-particle trajectories all of which have
shadows that last several tens of crossing times, although it is
surprising that the shadow durations are slightly {\em longer} than
that predicted by superimposing single-particle trajectories.
Now, if particles encounter glitches largely independently of one another,
and shadow durations for softened systems are long for most particles,
then we can hypothesize that reasonable statistical results may be
acquired from long simulations of large softened systems as long as only
a small number of particles have undergone glitches, and the statistics
taken depend on large numbers of particles.  Thus, for
example, the global spatial distribution of matter in a simulated
galaxy may be correct, but the number of escapers from a
simulated globular cluster may be incorrect if the stars that escape
happen also to be the stars that underwent glitches before escaping.

The difference between the shadow durations for softened \vs unsoftened
systems undoubtedly is related to fluctuating Lyapunov exponents as
discussed in
\cite{DawsonGrebogiSauerYorke94,SauerGrebogiYorke97,Sauer02}.
Although we have not measured Lyapunov exponents explicitly, we measured
a related quantity, namely the expansion and contraction factors across
a timestep of the vectors that span the locally expanding and contracting
spaces.  In a
uniformly hyperbolic system, these factors will always be greater and
less than 1, respectively.  An event of ``non-hyperbolicity'' can be
observed by looking for areas along the trajectory where directions
which were previously expanding over long periods instead start to
contract, or vice versa.
If we plot the expansion and contraction amounts
along a trajectory, we find that ``non-hyperbolic'' events correllate well
with the occurence glitches \footnote{Note that we did not
observe any obvious neutral directions, which is moderately surprising
given that Hamiltonian systems are not hyperbolic.}.
If we plot an $M=1$ particle orbit in
three dimensions, we also observe that these events loosely correllate
with times when the particle's orbital geometry changes in an obvious way.
We postulate that the locally expanding
and contracting directions of a particle in the system are closely
related to the geometry of the particle's orbit, so that
changing the geometry of the orbit can cause these local vectors to
become inconsistent as time progresses.  In an unsoftened system, the
geometry of a particle's
orbit can be suddenly and violently changed by a close encounter.
In softened systems, however,
there is no precise, short-duration ``event'' which triggers
non-hyperbolicity; instead, the geometry of the orbit of a particle
changes slowly, so that many crossing times occur before
a glitch is likely.  This helps to explain Figs.
\ref{fig:soft0-M1-expo-1e-14} and \ref{fig:soft-M1-hist}.

In conclusion, we postulate that there is no feasible integration
accuracy which will produce long shadows in unsoftened gravitational
$n$-body simulations.  This does not necessarily mean that such
simulations should not be trusted, only that shadowing may be too
stringent a measure of error.  In contrast, we believe that there {\em
does} exist a feasible integration accuracy for which softened systems
are shadowable for many crossing times even for large $n$.  Given the
stark dependence of shadow duration on the softening parameter, however,
we suspect that until a better understanding is acquired, the
shadowability of physical simulations in general will need to be
decided on a case-by-case basis. 

The author's software \cite{HayesMSc} is available on request.

%
%
%
%
%
%
%
%

{\bf Acknowledgments}:
We thank Scott Tremaine, Ken Jackson, and James Yorke for thoughtful
discussions on this work, and the University of Toronto's Computing
Disciplines Facility for cluster computer time.
This work was supported in part by the Natural Sciences and Engineering
Research Council of Canada, and the Information Technology Research
Centre of Ontario.

\bibliography{aps}

\begin{thebibliography}{18}
\expandafter\ifx\csname natexlab\endcsname\relax\def\natexlab#1{#1}\fi
\expandafter\ifx\csname bibnamefont\endcsname\relax
  \def\bibnamefont#1{#1}\fi
\expandafter\ifx\csname bibfnamefont\endcsname\relax
  \def\bibfnamefont#1{#1}\fi
\expandafter\ifx\csname citenamefont\endcsname\relax
  \def\citenamefont#1{#1}\fi
\expandafter\ifx\csname url\endcsname\relax
  \def\url#1{\texttt{#1}}\fi
\expandafter\ifx\csname urlprefix\endcsname\relax\def\urlprefix{URL }\fi
\providecommand{\bibinfo}[2]{#2}
\providecommand{\eprint}[2][]{\url{#2}}

\bibitem[{\citenamefont{Dawson et~al.}(3 Oct 1994)\citenamefont{Dawson,
  Grebogi, Sauer, and Yorke}}]{DawsonGrebogiSauerYorke94}
\bibinfo{author}{\bibfnamefont{S.}~\bibnamefont{Dawson}},
  \bibinfo{author}{\bibfnamefont{C.}~\bibnamefont{Grebogi}},
  \bibinfo{author}{\bibfnamefont{T.}~\bibnamefont{Sauer}}, \bibnamefont{and}
  \bibinfo{author}{\bibfnamefont{J.~A.} \bibnamefont{Yorke}},
  \bibinfo{journal}{Physical Review Letters} \textbf{\bibinfo{volume}{73}},
  \bibinfo{pages}{1927} (\bibinfo{year}{3 Oct 1994}).

\bibitem[{\citenamefont{Sauer et~al.}(7 July 1997)\citenamefont{Sauer, Grebogi,
  and Yorke}}]{SauerGrebogiYorke97}
\bibinfo{author}{\bibfnamefont{T.}~\bibnamefont{Sauer}},
  \bibinfo{author}{\bibfnamefont{C.}~\bibnamefont{Grebogi}}, \bibnamefont{and}
  \bibinfo{author}{\bibfnamefont{J.~A.} \bibnamefont{Yorke}},
  \bibinfo{journal}{Physical Review Letters} \textbf{\bibinfo{volume}{79}},
  \bibinfo{pages}{59} (\bibinfo{year}{7 July 1997}).

\bibitem[{\citenamefont{Sauer}(2002)}]{Sauer02}
\bibinfo{author}{\bibfnamefont{T.~D.} \bibnamefont{Sauer}},
  \bibinfo{journal}{Physical Review E} \textbf{\bibinfo{volume}{65}},
  \bibinfo{pages}{036220} (\bibinfo{year}{2002}).

\bibitem[{\citenamefont{Farmer and Sidorowich}(1991)}]{FarmerSidorowich91}
\bibinfo{author}{\bibfnamefont{J.~D.} \bibnamefont{Farmer}} \bibnamefont{and}
  \bibinfo{author}{\bibfnamefont{J.~J.} \bibnamefont{Sidorowich}},
  \bibinfo{journal}{Physica D} \textbf{\bibinfo{volume}{47}},
  \bibinfo{pages}{373} (\bibinfo{year}{1991}).

\bibitem[{\citenamefont{Fryska and Zohdy}(1992)}]{FryskaZohdy92}
\bibinfo{author}{\bibfnamefont{S.~T.} \bibnamefont{Fryska}} \bibnamefont{and}
  \bibinfo{author}{\bibfnamefont{M.~A.} \bibnamefont{Zohdy}},
  \bibinfo{journal}{Physics Letters A} \textbf{\bibinfo{volume}{166}},
  \bibinfo{pages}{340} (\bibinfo{year}{1992}).

\bibitem[{\citenamefont{Corless}(1994)}]{CorlessWhatGood94}
\bibinfo{author}{\bibfnamefont{R.~M.} \bibnamefont{Corless}},
  \bibinfo{journal}{Computers Math. Applic.} \textbf{\bibinfo{volume}{28}},
  \bibinfo{pages}{107} (\bibinfo{year}{1994}).

\bibitem[{\citenamefont{Anosov}(1967)}]{Anosov}
\bibinfo{author}{\bibfnamefont{D.~V.} \bibnamefont{Anosov}},
  \bibinfo{journal}{Proc. Steklov Inst. Math} \textbf{\bibinfo{volume}{90}},
  \bibinfo{pages}{1} (\bibinfo{year}{1967}).

\bibitem[{\citenamefont{Bowen}(1975)}]{Bowen}
\bibinfo{author}{\bibfnamefont{R.}~\bibnamefont{Bowen}},
  \bibinfo{journal}{Journal of Differential Equations}
  \textbf{\bibinfo{volume}{18}}, \bibinfo{pages}{333} (\bibinfo{year}{1975}).

\bibitem[{\citenamefont{Grebogi et~al.}(1990)\citenamefont{Grebogi, Hammel,
  Yorke, and Sauer}}]{GHYS}
\bibinfo{author}{\bibfnamefont{C.}~\bibnamefont{Grebogi}},
  \bibinfo{author}{\bibfnamefont{S.~M.} \bibnamefont{Hammel}},
  \bibinfo{author}{\bibfnamefont{J.~A.} \bibnamefont{Yorke}}, \bibnamefont{and}
  \bibinfo{author}{\bibfnamefont{T.}~\bibnamefont{Sauer}},
  \bibinfo{journal}{Physical Review Letters} \textbf{\bibinfo{volume}{65}},
  \bibinfo{pages}{1527} (\bibinfo{year}{1990}).

\bibitem[{\citenamefont{Quinlan and Tremaine}(1992)}]{QT}
\bibinfo{author}{\bibfnamefont{G.~D.} \bibnamefont{Quinlan}} \bibnamefont{and}
  \bibinfo{author}{\bibfnamefont{S.}~\bibnamefont{Tremaine}},
  \bibinfo{journal}{Monthly Notices of the Royal Astronomical Society}
  \textbf{\bibinfo{volume}{259}}, \bibinfo{pages}{505} (\bibinfo{year}{1992}).

\bibitem[{\citenamefont{Coomes et~al.}(1995)\citenamefont{Coomes, Ko{\c c}ak,
  and Palmer}}]{CoomesKocakPalmer95a}
\bibinfo{author}{\bibfnamefont{B.~A.} \bibnamefont{Coomes}},
  \bibinfo{author}{\bibfnamefont{H.}~\bibnamefont{Ko{\c c}ak}},
  \bibnamefont{and} \bibinfo{author}{\bibfnamefont{K.~J.}
  \bibnamefont{Palmer}}, \bibinfo{journal}{Numerische Mathematik}
  \textbf{\bibinfo{volume}{69}}, \bibinfo{pages}{401} (\bibinfo{year}{1995}).

\bibitem[{\citenamefont{Hayes}(2001)}]{HayesPhD}
\bibinfo{author}{\bibfnamefont{W.~B.} \bibnamefont{Hayes}}, Ph.D. thesis,
  \bibinfo{school}{Department of Computer Science, University of Toronto}
  (\bibinfo{year}{2001}).

\bibitem[{\citenamefont{Lai et~al.}(1999)\citenamefont{Lai, Lerner, Williams,
  and Grebogi}}]{LaiLernerWiliamsGrebogi99}
\bibinfo{author}{\bibfnamefont{Y.-C.} \bibnamefont{Lai}},
  \bibinfo{author}{\bibfnamefont{D.}~\bibnamefont{Lerner}},
  \bibinfo{author}{\bibfnamefont{K.}~\bibnamefont{Williams}}, \bibnamefont{and}
  \bibinfo{author}{\bibfnamefont{C.}~\bibnamefont{Grebogi}},
  \bibinfo{journal}{Physical Review E} \textbf{\bibinfo{volume}{60}},
  \bibinfo{pages}{5445} (\bibinfo{year}{1999}).

\bibitem[{\citenamefont{Heggie and Mathieu}(1986)}]{HeggieMathieuNoXRef}
\bibinfo{author}{\bibfnamefont{D.~C.} \bibnamefont{Heggie}} \bibnamefont{and}
  \bibinfo{author}{\bibfnamefont{R.~D.} \bibnamefont{Mathieu}}, in
  \emph{\bibinfo{booktitle}{{The Use of Supercomputers in Stellar Dynamics}}},
  edited by \bibinfo{editor}{\bibfnamefont{P.}~\bibnamefont{Hut}}
  \bibnamefont{and} \bibinfo{editor}{\bibfnamefont{S.~L.~W.}
  \bibnamefont{McMillan}} (\bibinfo{publisher}{Springer-Verlag},
  \bibinfo{year}{1986}), pp. \bibinfo{pages}{233--235}.

\bibitem[{\citenamefont{Hindmarsh}(1980)}]{LSODE}
\bibinfo{author}{\bibfnamefont{A.~C.} \bibnamefont{Hindmarsh}},
  \bibinfo{journal}{{ACM-SIGNUM Newsletter}} \textbf{\bibinfo{volume}{15}},
  \bibinfo{pages}{10} (\bibinfo{year}{1980}).

\bibitem[{\citenamefont{Spitzer~Jr.}(1987)}]{SpitzerGlob}
\bibinfo{author}{\bibfnamefont{L.}~\bibnamefont{Spitzer~Jr.}},
  \emph{\bibinfo{title}{{Dynamical Evolution of Globular Clusters}}}, Princeton
  series in astrophysics (\bibinfo{publisher}{Princeton University Press},
  \bibinfo{year}{1987}).

\bibitem[{\citenamefont{Hayes}(1995)}]{HayesMSc}
\bibinfo{author}{\bibfnamefont{W.}~\bibnamefont{Hayes}}, Master's thesis,
  \bibinfo{school}{Dept. of Computer Science, University of Toronto}
  (\bibinfo{year}{1995}).

\bibitem[{\citenamefont{Wisdom and Holman}(1992)}]{WisdomHolman92}
\bibinfo{author}{\bibfnamefont{J.}~\bibnamefont{Wisdom}} \bibnamefont{and}
  \bibinfo{author}{\bibfnamefont{M.}~\bibnamefont{Holman}},
  \bibinfo{journal}{The Astronomical Journal} \textbf{\bibinfo{volume}{104}},
  \bibinfo{pages}{2022} (\bibinfo{year}{1992}).

\end{thebibliography}

\end{document}